\documentclass[aps]{revtex4}
\usepackage[colorlinks=true, pdfstartview=FitV, linkcolor=blue, citecolor=red, urlcolor=magenta, breaklinks=true]{hyperref}
\usepackage{graphicx}  
\usepackage{amsfonts}
\usepackage{float}
\usepackage{amssymb}
\newcommand{\sech}{{\,\rm sech}}

\begin{document}


\title{Kondo effect  from a Lorentz-violating domain wall description of superconductivity}

\author{D. Bazeia$^{1}$, F.A. Brito$^{1,2}$, J.C. Mota-Silva$^{1}$}
\email{dbazeia@gmail.com, fabrito@df.ufcg.edu.br, julio$_$cmota@yahoo.com.br}
\address{$\,^{1}$ Departamento de F\'isica, Universidade Federal da Para\'iba, Caixa Postal 5008, 58051-970 Jo\~ao Pessoa, Para\'iba, Brazil\\ $\,^{2}$ 
Departamento de F\'isica, Universidade Federal de Campina Grande, Caixa Postal 10071,58109-970 Campina Grande, Para\'iba, Brazil.}

\begin{abstract}
We extend recent results on domain wall description of superconductivity in an Abelian Higgs model by introducing a particular Lorentz-violating term.  The temperature of the system is interpreted through the fact that the soliton following accelerating orbits is a Rindler observer experiencing a thermal bath. We show that this term can be associated with the {\sl Kondo effect}, that is, the Lorentz-violating parameter is closely related to the concentration of magnetic impurities living on a superconducting domain wall. We also found that the critical temperature decreasing with the impurity concentration as a non single-valued function,  for the case $T_K < T_{c0}$, develops a negative curvature and presents deviations from the Abrikosov and Gor'kov theory, a phenomenon  already supported by experimental evidence.
\end{abstract}
\maketitle
\pretolerance10000

\section{Introduction}
In a recent study was put forward an alternative theory of superconductivity in field theory through domain wall description of superconductivity \cite{eu}. Now we extend this investigation by considering a theory that allows Lorentz and CPT violation. The possibility of breaking Lorentz and CPT symmetries has been considered in many works in the literature \cite{carrolljackiw, Kostelecky,cpt}. On the other hand, 
there are many studies on superconducting solitons as, for example, superconducting strings \cite{vilenkin, witten} and related solutions such as domain walls with internal structures \cite{paredes1, paredes2}. Since such soliton solutions can follow non-trivial orbits in the field space \cite{paredes2}, they are mostly forced to move along accelerated trajectories. In the limit of small velocities these solutions can be identified as Rindler observers  \cite{eu} in a thermal bath which allows to introduce temperature in the system via Unruh effect, supporting an investigation of thermodynamic quantities in the domain wall description of superconductivity. 
The quantum field theory can explain several important effects of superconductivity through an appropriate classical regime of a quantum field theory inspired by the Ginzburg-Landau (GL) theory \cite{livro, GL}. In the present model, this description uses a dynamical complex scalar field coupled to Abelian gauge field (the Abelian Higgs sector) which is responsible for superconductivity of the system and an extra real scalar field responsible for the domain walls plus other terms that breaks the Lorentz and CPT symmetries. 

Lorentz and CPT symmetries are important properties in particle physics models and the possibility of breaking these symmetries has been considered in several different contexts \cite{carrolljackiw, Kostelecky}. The models considering Lorentz and CPT symmetries violations as extensions of the standard model can modify the scalar Higgs sector and this gives room for defect structures of more general profiles \cite{cpt}.

The main  feature presented in our model is that it can describe a domain wall superconductivity whose Lorentz-violating term can play the role of magnetic impurities in the superconductor. Magnetic impurities have a number of striking effects on superconductivity and one of their effects is the Kondo effect \cite{muller,tami}. As we shall show,  our model can describe the well-known competition between Kondo effect and superconductivity \cite{braz,science} at the domain wall. \\
\indent The Letter is organized as follows. In Sec.~\ref{1} we present the model enriched with the term that allows the Lorentz and CPT symmetries breaking, which enjoys a new parameter --- the {\it Lorentz-violating parameter}. In Sec.~\ref{2} we consider the soliton solutions as background fields to solve a Schroedinger-like equation for the electromagnetic field. In Sec.~\ref{IV} we calculate the condensate at finite temperature as a function of the new parameter. In Sec.~\ref{V} we find the optical conductivity of the system and discuss the influence of the new parameter. In Sec.~\ref{kondo}, we establish a clear relationship between the Lorentz-violating parameter and the concentration of impurities of a superconducting material with magnetic impurities which develops the {\sl Kondo effect}. Finally, in the Sec.~\ref{conclu} we make our conclusions. 
 
\section{The model with $\kappa^{\mu\nu}$.}
\label{1}
In this section we follow the Lagrangian that describes the superconducting domain wall given by \cite{eu}
\begin{eqnarray}
{\cal L} = \frac{1}{2}\partial_{\mu}\phi\partial^{\mu}\phi + (\partial^{\mu}\chi + iqA^{\mu}\chi)(\partial_{\mu}\chi^{*} - iqA_{\mu}\chi^{*}) -V(\phi,\chi,\chi^{*}) - \frac{1}{4}F_{\mu\nu}F^{\mu\nu} \label{0},
\end{eqnarray}
where $\mu,\nu=0,1,2,...,d$ are bulk indices for arbitrary $(d-2)$-dimensional domain walls. In the present case, we shall focus on 2-dimensional domain walls such that $\mu,\nu=t,x,y,r$.

Let us now extend this Lagrangian to study a model with the possibility of Lorentz-symmetry violation. Firstly, we consider the class of models with real scalar fields  \cite{cpt}
\begin{eqnarray}
{\cal L} = \frac{1}{2}\partial_{\mu}\phi\partial^{\mu}\phi + \frac{1}{2}\kappa^{\mu\nu}\partial_{\mu}\phi\partial_{\nu}\phi + \frac{1}{2}\partial_{\mu}\chi\partial^{\mu}\chi + \frac{1}{2}\kappa^{\mu\nu}\partial_{\mu}\chi\partial_{\nu}\chi - V(\phi,\chi),
\label{1.2}
\end{eqnarray}
where $\kappa^{\mu\nu}$ is a constant tensor whose components {in general} read
\begin{equation}
\kappa^{\mu\nu}=
\left(\begin{array}{clcl}
\zeta &\varepsilon &\varepsilon & \varepsilon\\
\varepsilon &\zeta &\varepsilon &\varepsilon \\
\varepsilon &\varepsilon &\zeta &\varepsilon\\
\varepsilon &\varepsilon &\varepsilon &\zeta
\end{array}\right), 
\end{equation}
with $\zeta$ and $\varepsilon$ being real parameters and the scalar potential is given in terms of a `superpotential' $W$ as follows
\begin{eqnarray}
V(\phi,\chi)=\frac12W_\phi^2+\frac12W_\chi^2,
\end{eqnarray}
where  $W_{\phi_i}$ stands for partial derivatives of $W$ with respect to fields $\phi_i$.
The equations of motion are given by
\begin{eqnarray}
\Box\phi + \kappa^{\mu\nu}\partial_{\mu}\partial_{\nu}\phi = - \frac{\partial V(\phi,\chi)}{\partial \phi} 
\label{3},
\end{eqnarray}
\begin{eqnarray}
\Box\chi + \kappa^{\mu\nu}\partial_{\mu}\partial_{\nu}\chi = - \frac{\partial V(\phi,\chi)}{\partial \chi} 
\label{4}.
\end{eqnarray}
Now for simplicity, {but  keeping the scenario still rich enough for our present discussion}, we take $\varepsilon= 0$, 
then the equations of motion for the fields $\phi\equiv\phi(t,r)$ and $\chi\equiv\chi(t,r)$ are
\begin{eqnarray}
\ddot{\phi} - \phi'' + \zeta(\ddot{\phi} + \phi'') = - \frac{\partial V(\phi,\chi)}{\partial \phi},\\
\label{5} 
\ddot{\chi} - \chi'' + \zeta(\ddot{\chi} + \chi'') = - \frac{\partial V(\phi,\chi)}{\partial \chi}.
\label{5.1}
\end{eqnarray}
Thus, for static solutions we get
\begin{eqnarray}
\phi''(1 - \zeta) =  \frac{\partial V(\phi,\chi)}{\partial \phi},\\
\label{6} 
\chi''(1 - \zeta) =  \frac{\partial V(\phi,\chi)}{\partial \chi}.
\label{6.1}
\end{eqnarray}
By making the following transformation on the transversal coordinate $r$
\begin{eqnarray}
\widetilde{r} = \frac{r}{\sqrt{1-\zeta}},
\end{eqnarray}
the equations of motion for the scalar fields can be rewritten as
\begin{eqnarray}
{\phi}''(\widetilde{r})  = \frac{\partial V}{\partial \phi}, 
\label{8}
\end{eqnarray}
and
\begin{eqnarray}
{\chi}''(\widetilde{r})  = \frac{\partial V}{\partial \chi}.
\label{9}
\end{eqnarray}

The dynamics of the model (\ref{1.2}) governed by the equations of motion (\ref{8}) and (\ref{9}) can be reduced to the first order equations  \cite{paredes1,paredes2}
\begin{eqnarray}
\frac{d\phi}{d\widetilde{r}}=W_\phi, \qquad \frac{d\chi}{d\widetilde{r}}=W_\chi.
\end{eqnarray}
For a specific superpotential 
\begin{eqnarray}
W(\phi,\chi)=\lambda\left(\frac13\phi^3-\phi\, a^2 \right)+\mu \phi\chi^2
\end{eqnarray}
the model produces domain wall solutions whose kink profiles are the following well-known BPS static solutions, known as type I solution
\begin{eqnarray}
{\phi}&=&-a\tanh(\lambda a\widetilde{r})\nonumber\\
{\chi}&=&0
\label{10}
\end{eqnarray}
and type II solution
\begin{eqnarray}
{\phi}&=&-a\tanh(2\mu a\widetilde{r})\nonumber\\
{\chi}&=&\pm a\sqrt{\frac{\lambda}{\mu}-2} \sech(2\mu a\widetilde{r}).
\label{11}
\end{eqnarray}
For latter convenience, the type II solution in terms of the original coordinate $r$ is
\begin{eqnarray}
\widetilde{\phi}&=&-a\tanh\left(\frac{2\mu a}{\sqrt{1-\zeta}}r\right)
\nonumber\\
\widetilde{\chi}&=&\pm a\sqrt{\frac{\lambda}{\mu}-2} \sech\left(\frac{2\mu a}{\sqrt{1-\zeta}}r\right).
\label{13}
\end{eqnarray}
\section{Superconducting Type II domain wall solution}
\label{2}
The superconducting domain wall developing a condensate can be obtained from the following Lagrangian with Lorentz-violating symmetry but preserves the $Z_{2}\times U(1)$ symmetry: 
\begin{eqnarray}
{\cal L} &=& \frac{1}{2}\partial_{\mu}\phi\partial^{\mu}\phi + \frac{1}{2}\kappa^{\mu\nu}\partial_{\mu}\phi\partial_{\nu}\phi + D_{\mu}\chi (D^{\mu}\chi)^{*} + \kappa^{\mu\nu}D_{\mu}\chi (D_{\nu}\chi)^{*}\nonumber\\ &-& V(\phi,\chi, \chi^{*}) - \frac{1}{4}F_{\mu\nu}F^{\mu\nu},
\label{14}
\end{eqnarray}
where $D_{\mu}= \partial_{\mu} - iqA_{\mu}$ and $F^{\mu\nu} = \partial^{\mu}A^{\nu}-\partial^{\nu}A^{\mu}$. The potential is appropriately chosen as 
\begin{eqnarray}
V(\phi,\chi, \chi^{*}) = \frac{1}{2}\lambda^{2}(\phi^{2} - a^{2})^{2} + \lambda\mu(\phi^{2} - a^{2})\left|\chi\right|^{2} + \frac{1}{2}\mu^{2}\left|\chi\right|^{4} + \mu^{2}\phi^{2}\left|\chi\right|^{2}.
\label{15}
\end{eqnarray}
The equations of motion for the coupled complex scalar and electromagnetic fields are given by 
\begin{eqnarray}
&&\Box\chi + \kappa^{\mu\nu}[\partial_{\mu}\partial_{\nu}\chi - iqA^{\nu}\partial_{\nu}\chi - iqA_{\mu}\partial_{\mu}\chi - q^{2}A_{\mu}A_{\nu}\chi]\nonumber\\ &&+ \frac{\partial V}{\partial\chi^{*}} - q^{2}A_{\mu}A^{\nu}\chi - 2iqA^{\gamma}\partial_{\gamma}\chi = 0,
\label{16}
\end{eqnarray}
\begin{eqnarray}
&&\Box A^{\theta} +\kappa^{\mu\nu}[iq\chi^{*}\partial^{\theta}\chi - iq\chi\partial^{\theta}\chi^{*} + q^{2}\delta^{\theta}_{\mu}A_{\nu}\left|\chi\right|^{2} + q^{2}\delta^{\theta}_{\nu}A_{\mu}\left|\chi\right|^{2}] + \nonumber\\ &&+ iq[\chi^{*}\partial^{\theta}\chi - \chi\partial^{\theta}\chi^{*}]- 2q^{2}A^{\theta}\left|\chi\right|^{2}= 0. 
\label{17}
\end{eqnarray}
Now considering the equations (\ref{13}) as background field solutions we solve the equation (\ref{17}) by introducing
\begin{eqnarray}
A_{\mu}(t,r) = A_{\mu}(r)\exp{(-i\omega t)}\qquad \chi(t,r) = \widetilde{\chi}(r)\exp{(-i\theta t)},
\label{20}
\end{eqnarray}
such that we get
\begin{eqnarray}
-A_{x}''+2(1-\zeta)q^{2}A_{x}\left|\widetilde{\chi}\right|^{2} = \omega^{2}A_{x}.
\label{21}
\end{eqnarray}
Let us make the following redefinitions
\begin{eqnarray}
\alpha \equiv \frac{2\mu a}{\sqrt{1-\zeta}}, \qquad \frac{1}{4}\ell^{2} = 2\tilde{q}^{2}a^{2}\left(\frac{\lambda}{\mu} - 2\right), \qquad \tilde{q} = \sqrt{1-\zeta}q.
\label{22}
\end{eqnarray}
Thus,
\begin{eqnarray}
-A_{x}'' + \frac{1}{4}\ell^{2}\sech^{2}(\alpha r)A_{x} = \omega^{2}A_{x},
\label{23}
\end{eqnarray}
which solution is
\begin{eqnarray}
A_{x}(\omega,\alpha,\ell,r) = (\sech(\alpha r))^{\frac{i\omega}{\alpha}}\,_{2}F_{1}\left[a1;a2;a3;\frac{1}{2}(1-\tanh(\alpha r))\right],
\label{24}
\end{eqnarray}
where the parameters $a1, a2$ and $a3$ are given by
\begin{eqnarray}
a1 = \frac{1}{2}\frac{-2i\omega + \alpha + \sqrt{-\ell^{2}+ \alpha^{2}}}{\alpha},
\label{25}
\end{eqnarray}
\begin{eqnarray}
a2 = -\frac{1}{2}\frac{2i\omega - \alpha + \sqrt{-\ell^{2}+ \alpha^{2}}}{\alpha},
\label{26}
\end{eqnarray}
\begin{eqnarray}
a3 = -\frac{i\omega - \alpha}{\alpha}.
\label{27}
\end{eqnarray}
\section{The condensate at finite temperature}
\label{IV}
To introduce the effects of temperature in the system we identify the solutions $\widetilde{\phi}$ and $\widetilde{\chi}$ in terms of an accelerated observer coordinates as follows \cite{eu}
\begin{eqnarray}
\widetilde{\phi} = \alpha a t(\tau),
\label{28}
\end{eqnarray}
\begin{eqnarray}
\widetilde{\chi} = \alpha\left(\frac{\lambda}{\mu}-2\right)^{\frac{1}{2}} a z(\tau),
\label{29}
\end{eqnarray} 
where
\begin{eqnarray}
t(\tau) = \frac{1}{\alpha}\tanh{\alpha\tau},
\label{30}
\end{eqnarray}
\begin{eqnarray}
z(\tau) = \frac{1}{\alpha}\sech{\alpha\tau}.
\label{31}
\end{eqnarray} 
Note that $(\tau)$ is identified wich Euclidean time. We use the definition of the acceleration to establish the following relationship
\begin{eqnarray}
a_c &\equiv& \sqrt{a_{\mu}a^{\mu}} = \alpha - \frac{1}{2}\alpha(\alpha\tau)^{2}+...\nonumber\\&\approx&\alpha.
\label{32}
\end{eqnarray} 
In this approximation (regime of small velocities) the accelerated observer is of the Rindler type.
We use the Unruh temperature to find ${\alpha}\approx2\pi \widetilde{T}$, where ({up to a numerical factor of $2\pi$ that can be absorbed into $\tau$}) we shall simply identify the modified temperature as
\begin{eqnarray}
{\alpha}  \equiv \widetilde{T}=\frac{2\mu a}{\sqrt{1-\zeta}}=\frac{T}{\sqrt{1-\zeta}}.
\label{33}
\end{eqnarray}
{This temperature obtained using the type II solution is the temperature inside the condensate --- see below. For temperatures $T\geq T_c$, where $T_c$ is the critical temperature we take advantage of the type I solution (\ref{10}). Thus, following Eq.~(\ref{33}) is natural to define the modified critical temperature in terms of the energy scale $\lambda a$ of (\ref{10}) as follows
\begin{eqnarray}
\widetilde{T}_c=\frac{\lambda a}{\sqrt{1-\zeta}}= \frac{T_c}{\sqrt{1-\zeta}}.
\label{33c}
\end{eqnarray}
} 
Since the aforementioned type II BPS solution follows elliptic orbits \cite{paredes1,paredes2,eu}, then we can also define the acceleration of the system (in the target space) as
\begin{eqnarray}
\left.\frac{d^{2}\widetilde{\chi}}{d\widetilde{\phi}^{2}}\right|_{{r}_0}&=&\left.\frac{d}{d\widetilde{\phi}}\left(\frac{d\widetilde{\chi}}{d\widetilde{\phi}}\right)\right|_{{r}_0} = \left.\frac{d}{d\phi}\left(\frac{d W_{\widetilde{\chi}}}{d W_{\widetilde{\phi}}}\right)\right|_{{r}_0} = \frac{W_{\widetilde{\phi}\widetilde{\chi}}}{W_{\widetilde{\phi}}} 
- \left.\frac{W_{\widetilde{\phi}\widetilde{\phi}}W_{\widetilde{\chi}}}{W^{2}_{\widetilde{\phi}}}\right|_{{r}_{0}}\nonumber\\&=&-\frac{\sqrt{\frac{\lambda}{\mu}-2}}{a\sech(\alpha{r}_{0})}\left(1+\frac{\lambda}{\mu}\frac{\tanh^{2}(\alpha {r}_{0})}{\sech^{2}(\alpha {r}_{0})}\right).\label{34}
\end{eqnarray}
Since the coordinates $(\widetilde{\phi},\widetilde{\chi})$ has dimension of energy, the acceleration in the target space can be now related to the (Unruh) temperature of the system as in the form \cite{eu}
\begin{eqnarray}
\beta = \left|\frac{d^{2}\widetilde{\chi}}{d\widetilde{\phi}^{2}}\right|_{{r}_{0}}, \qquad \beta = \frac{1}{T}.
\label{35}
\end{eqnarray}
{In order to support the above statement, notice that from equations (\ref{28})-(\ref{31}) we find 
\begin{eqnarray}
\frac{d^{2}\widetilde{\chi}}{d\widetilde{\phi}^{2}}=\frac{1}{\alpha}\frac{d^{2}\bar{z}}{d\bar{t}^{2}} = \frac{\alpha_0}{\alpha}=\frac{1}{T},
\label{ac-beta}
\end{eqnarray}
where $\alpha_0$ is identified with a dimensionless constant acceleration. The coordinates $\bar{z}=a\left(\frac{\lambda}{\mu}-2\right)^{1/2}z(\tau)$  and $\bar{t}=at(\tau)$ are dimensionless. In Eq.~(\ref{ac-beta}) it was assumed $\alpha_0=2\pi/\sqrt{1-\zeta}$ and $\alpha=2\pi \widetilde{T}$  (here the $2\pi$ factor is restored).
}

\begin{figure}[h!]
  \centering
    \includegraphics[width=8.0cm,height=5.5cm]{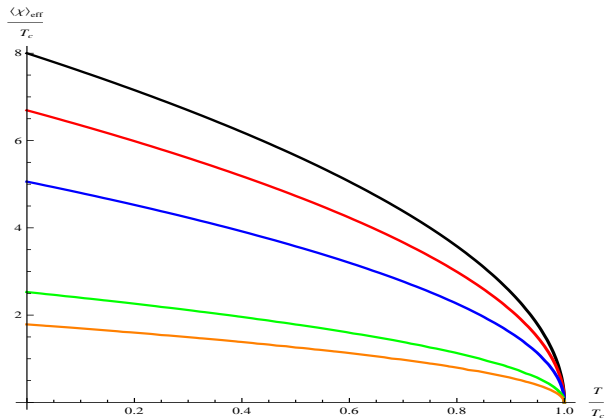}
    \caption{The effective condensate as a function of temperature for charge $q = 2$ and $\zeta$ = 0, 0.3, 0.6, 0.9 and 0.95 from top to bottom.}
    \label{figura2}
\end{figure}

We now expand the scalar solution $\widetilde{\chi}({r})$ around the core of the type II domain wall at ${r}\approx 0$ that is 
\begin{eqnarray}
\widetilde{\chi}({r}) = m - \frac{1}{2}m\alpha^{2}{r}^2 + ...
\end{eqnarray}
to isolate the condensate defined as $\left\langle \widetilde{\chi}\right\rangle \approx m$.  Now from Eqs.~(\ref{33})-(\ref{33c}) the temperature dependence is
\begin{eqnarray}
m = a\sqrt{\frac{\lambda}{\mu} - 2}\equiv a\sqrt{2}\sqrt{\frac{T_c}{T}-1}.
\end{eqnarray}
In order to obtain $a$ we can use Eqs.~(\ref{34})-(\ref{35}) in the `supersymmetric vacuum' ($\widetilde{\phi} = 0$ and $\widetilde{\chi} = \pm a\sqrt{\frac{\lambda}{\mu}}$), that is, in the limit ${r}_{0}\approx 0$ and $\frac{\lambda}{\mu}\gg1$, which as a function of the temperature reads
\begin{eqnarray}
\frac{1}{T}\approx \frac{\sqrt{\frac{T_c}{T}}}{a}\to a\sim T_c^{1/2}T^{1/2},
\label{36}
\end{eqnarray}
thus,
\begin{eqnarray}
m = \sqrt{2}T_{c}\sqrt{1 - \frac{T}{T_{c}}}.
\label{m}
\end{eqnarray}
This implies that the condensate has precisely the desired  form $\left\langle \widetilde{\chi}\right\rangle \approx m = \sqrt{2}T_{c}\sqrt{1-\frac{T}{T_{c}}}$. The {\it effective condensate} which is felt by the electromagnetic field through Eq.~(\ref{21}) is $\left\langle \widetilde{\chi}\right\rangle_{eff} \approx 4\tilde{q}\,T_{c}\sqrt{1 - \frac{T}{T_{c}}}$ with dependence on the {\it screened} electric charge $\widetilde{q}=\sqrt{1-\zeta}q$. See Fig.~\ref{figura2}.

\section{The conductivity and impurities}
\label{V}
As well-known, from the Ohm's law one can readily obtain the conductivity along a direction, say $x$-direction along
the domain walls, in the form
\begin{eqnarray}
\sigma_{x}(x,y) = \frac{J_{x}}{E_{x}} = \frac{A'_{x}(0)}{i\omega A_{x}(0)},
\label{c1}
\end{eqnarray}
where in the last step we have considered the fact that $E_x=-\partial_t A_x=i\omega A_x$ and defined the current density as $J_x=A'_x(0)$. The latter can be easily justified by using the boundary conditions for the electromagnetic field on an interface at some fixed point ($r=0$ or arbitrary point) of the transversal coordinate with the presence of a surface current. See Ref.~\cite{eu}, for further details on this issue.

\begin{figure}[h!]
  \centering
    \includegraphics[width=8.0cm,height=5.5cm]{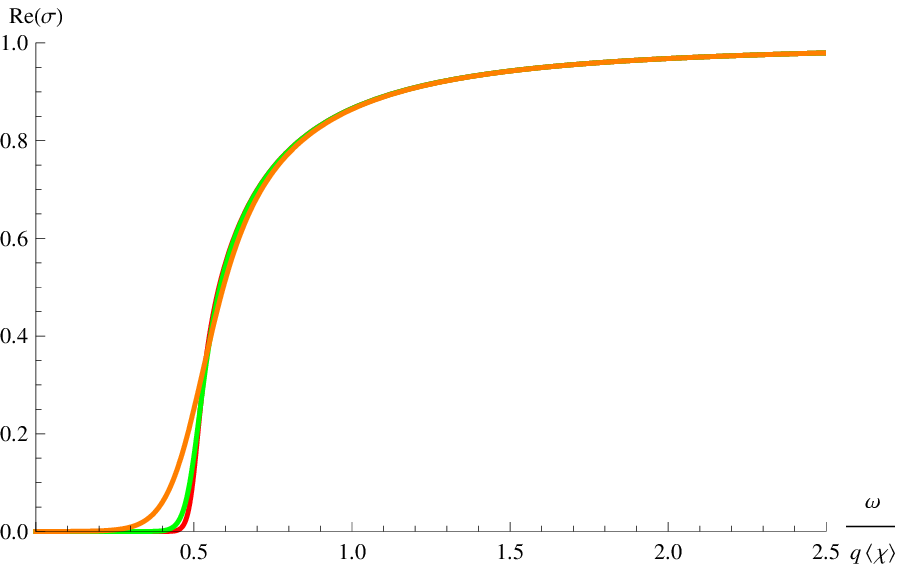}
    \includegraphics[width=8.0cm,height=5.5cm]{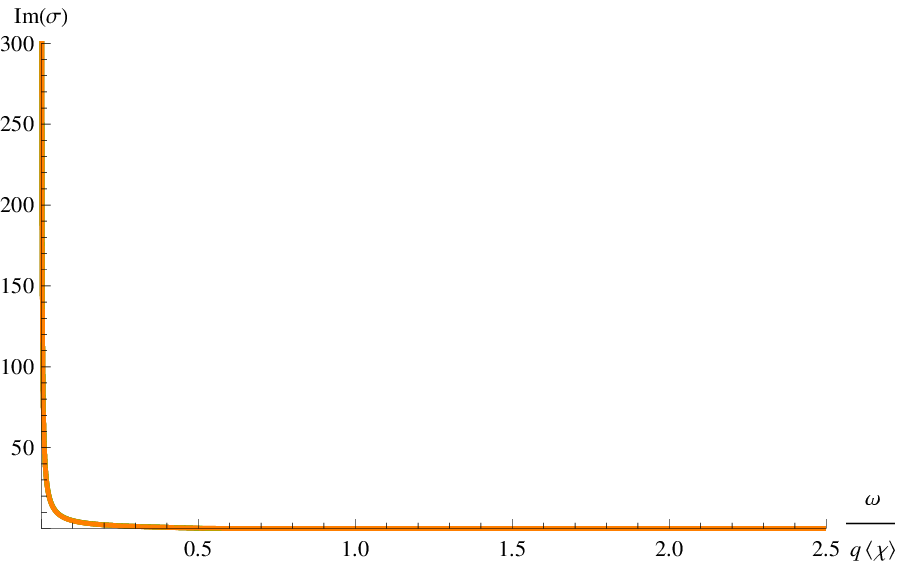}
    \caption{The real (left panel) and imaginary (right panel) part of conductivity as a function of the frequency normalized by the effective condensate for screened  charges $\tilde{q}$ at $\zeta =0.937500, 0.609375$ and 0 from top to bottom; $\delta = 0$ and $q=32$.} 
    \label{figura3}
\end{figure}

Now using our solution (\ref{24}) for the electromagnetic field around a generic point $r\simeq\delta$ we can write the conductivity as follows
\begin{eqnarray}
\sigma_{x}(\omega,\alpha,\ell,\delta) &=& \frac{\frac{1}{8}i(4\omega^{2} + 4i\omega\alpha-\ell^{2})\,_{2}F_{1}[b1,b2;b3;\frac{1}{2}(1-\tanh{(\alpha\delta)})]\sech{(\delta\alpha)}^{2}}{\omega(i\omega-\alpha)\,_{2}F_{1}[a1,a2;a3;\frac{1}{2}(1-\tanh(\alpha\delta))]} \nonumber\\&+& \tanh(\alpha\delta),
\label{c2}
\end{eqnarray}
where the parameters $b1$, $b2$ and $b3$ are defined by
\begin{eqnarray}b1 = -\frac{1}{2}\frac{2i\omega - 3\alpha + \sqrt{-\ell^{2} + \alpha^{2}}}{\alpha},\end{eqnarray}
\begin{eqnarray}b2 = -\frac{1}{2}\frac{-2i\omega + 3\alpha + \sqrt{-\ell^{2} + \alpha^{2}}}{\alpha},\end{eqnarray}
\begin{eqnarray}b3 = \frac{-i\omega - 2\alpha }{\alpha}\label{c3}.\end{eqnarray}
We consider the conductivity normalized by the effective condensate $\ell \rightarrow \tilde{q}\ell$, such that we define $\alpha = \tilde{q}^{-1}\tilde{q}\ell$ and $\omega=\omega_{r}\tilde{q}\ell$ into $\sigma$. The Fig.~\ref{figura3} shows the real and imaginary parts of the conductivity  as a function of $\frac{\omega}{\tilde{q}\left\langle \chi\right\rangle}$ (reduced frequency). 
The effect of values of the screened charges via influence of $\zeta$ can be noticed. This clearly shows a first evidence for impurities effect in the optical conductivity as a consequence of the Lorentz-violation term described by the Lorentz-violating parameter.  These effects in the optical conductivity remind those results obtained in a recent research about impurity effects in holographic superconductors \cite{hologra}.  We shall address further evidences of impurities being described in Lorentz-violating quantum field theory --- see below.
\begin{figure}[h!]
  \centering
    \includegraphics[width=8.0cm,height=5.5cm]{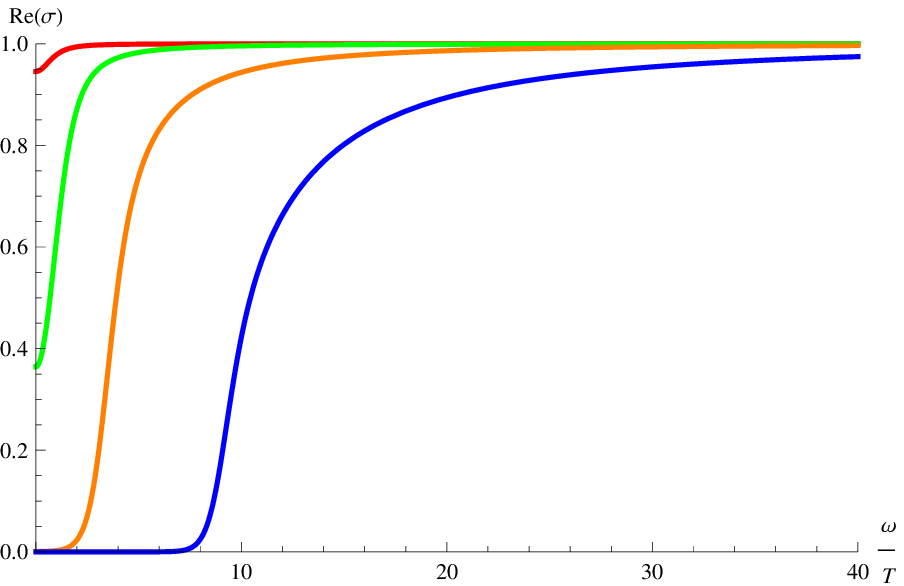}
    \includegraphics[width=8.0cm,height=5.5cm]{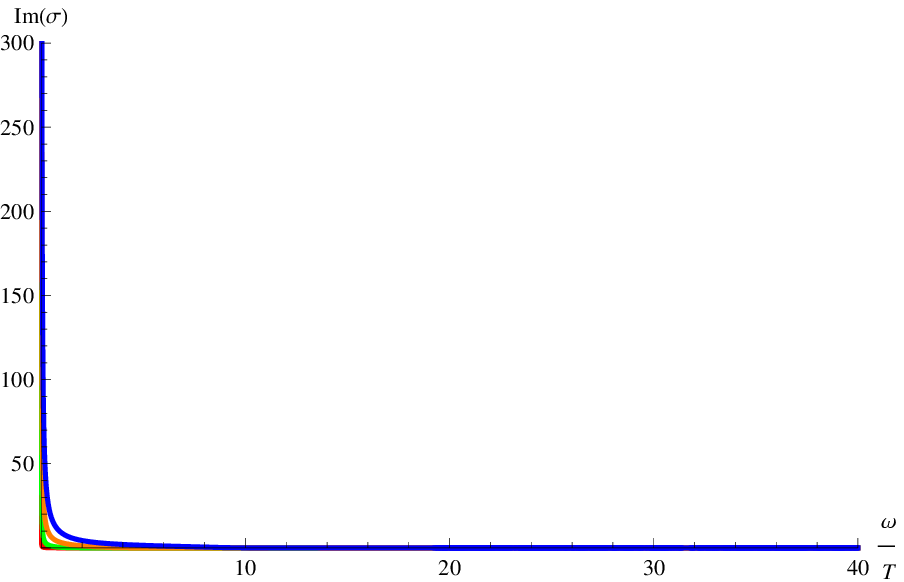}
    \caption{The real (left panel) and imaginary (right panel) part of conductivity as a function of the frequency normalized by the temperature for charge $q = 1$, $\zeta=0$ and temperatures $T = 0.99, 0.85, 0.45$ and $0.20$ from top to bottom; $\delta=0$ and the critical temperature is $T_c =1$.}
    \label{figura3T}
\end{figure}

The conductivity for $\zeta=0$ (with no impurities) as a function of the frequency normalized by the temperature can also be noticed in Fig.~\ref{figura3T}. The figures also show the presence of a pole in the imaginary part of the conductivity. Indeed there is always a pole in Im $(\sigma)$ as long as $\ell\equiv \left\langle \widetilde{\chi}\right\rangle_{eff}$ is real, i.e. as $T<T_c$, since one can be shown that 
\begin{equation}
\mbox{Im}\;\sigma(\omega)\to \frac{\ell}{\omega}, \qquad {\rm as} \quad \omega\to0.
\end{equation}
This can also be confirmed by noticing in Fig.~\ref{figura3T} that the more the temperature decreases a pole in the imaginary part is favored (blue line). As the temperature approaches the critical temperature (red line) the pole starts to disappear. Then according to the Kramers-Kronig relation 
\begin{equation}
\mbox{Im}\;\sigma(\omega)=-\frac{1}{\pi}{\cal P}\int_{-\infty}^\infty{\frac{\mbox{Re}\;\sigma(\omega')d\omega'}{\omega'-\omega}}
\end{equation}	
the existence of such a pole implies an existence of a delta function in $\mbox{Re}\;\sigma(\omega)$ at $\omega=0$ for temperatures below the critical temperature $T_c$. 
Thus our model presents an infinite DC conductivity as expected for a superconductor. 

\section{Kondo effect}
\label{kondo}

{ We notice from Fig.~\ref{figura2} that as $\zeta\to 1$ the screening charge $\tilde{q}$ goes to zero and in tuns $\ell\equiv \left\langle \widetilde{\chi}\right\rangle_{eff}\,\to 0$. At this regime the superconductivity is destroyed even at $T<T_c$. If we assume $\zeta$ as an external magnetic field, the existence of superconductivity imposes a limit on it. On the other hand, this may also be related to Kondo effect, where magnetic impurities decreases the conductivity. It is related to the increasing of the electric resistivity at low temperature as a response of the strong interaction between the conduction electrons and magnetic impurities in a normal metal or superconductor material. The main result that we shall present below seems to show a strong evidence of the Kondo effect (or at least a Kondo-like effect). Indeed the decreasing of the condensate as $\zeta\to1$ has particular resemblance with the pair-breaking due to magnetic impurities, reported long ago in \cite{muller,tami}. As we shall show explicitly, the critical temperature tends to decrease as a function of impurity concentration $\bar{n}\sim1/\rho$, where $\rho$ is the density of states of conduction electrons per atom per spin. In order to make a precise connection of our results with the Kondo effect we first expand the well-known critical temperature \cite{tami} in powers of $n$, i.e. 
\begin{eqnarray}
 T_c\sim T_K\exp{\left(-\frac{1}{|g|N\rho-\frac{n}{T_K\rho}}\right)} =T_{c0}\left(1-\frac{n}{T_Kc^2\rho}\right)+{\cal O}(n^2).
\label{33c-a}
\end{eqnarray}
This is for $T_{c0}\ll T_K<T_D$ and electrons with energy lesser than $T_K$.  The superconducting interaction $|g|$ is reduced by a repulsion $\sim n/NT_K\rho^2$ between impurity and electrons.
Here $c=|g|N\rho$ is depending on the superconducting interaction $g$, $\rho$ and the number of the atoms $N$. $T_{c0}\equiv T_Ke^{-1/c}$ is the critical temperature without impurities and $T_K$ is the Kondo temperature.  Eq.~(\ref{33c-a}) can be recast  in the form
\begin{eqnarray}
 \frac{T_c}{T_{c0}}=1-\frac{(2\pi)^2 T_{c0}}{T_Kc^2}\bar{n}+{\cal O}(n^2),
\label{33c-b}
\end{eqnarray}
where $\bar{n}=n/(2\pi)^2T_{c0}\rho$ is the impurity concentration. {This is the simplest regime in order to make a quicker and clear connection between the impurity concentration and the Lorentz-violating parameter $\zeta$ --- See below. In the following we shall address other interesting regimes.}
}

Interestingly, for large $\alpha\delta$ the argument in the hypergeometric functions in Eq.~(\ref{c2}) goes to zero as $e^{-2\alpha\delta}$. In this regime we can expand the conductivity formula by 
a power series of few terms given by
\begin{eqnarray}
\mbox{Re}\,\sigma(\omega,\alpha)\approx \delta(\omega)\left(1-\frac{1}{8}\frac{\ell^{2}}{\alpha^{2}}e^{-2\alpha\delta} + ...\right) \cong \delta(\omega)e^{-\frac{1}{8}\left(\frac{\Delta}{\alpha}\right)^{2}}
,\label{c4}
\end{eqnarray}
where
\begin{eqnarray}
\Delta =\ell e^{-\alpha\delta},
\label{c5}
\end{eqnarray}
{precisely defines the binding energy $\Delta(T\!=\!0)\!\simeq\!2T_c$ of a Cooper pair as long as we identify $\ell = 2\omega_D\equiv 2T_D$ as the Debye temperature and $\delta\alpha = 1/V N_F$ , being $V > 0$ the binding potential and $N_F$ the density of orbitals with Fermi's energy according to the Bardeen-Cooper-Schrieffer (BCS) theory of superconductors.
Thus, the critical temperature is defined as
\begin{eqnarray}
T_c = T_D e^{-\alpha{\delta}}.
\label{c5-a}
\end{eqnarray}
Since we are assuming the exponent in terms of a binding potential $V$ it might be clear that the parameter $\zeta$ indeed drives a variation of such binding potential as follows,
\begin{eqnarray}
\widetilde{V}=V\sqrt{1-\zeta}=V-\Delta V,
\label{c5-b}
\end{eqnarray}
where $\Delta V=(1/2)V\zeta+{\cal O}(\zeta^2)$ clearly weakens the binding energy of the Cooper pairs.
Then, recalling that $\alpha\equiv T/\sqrt{1-\zeta}$ and assuming the same previous reasoning the critical temperature becomes 
\begin{eqnarray}
T_c = T_D e^{-T\delta-\frac12T\delta\zeta}=T_{c0}e^{-\frac12T\delta\zeta}.
\label{c5-c}
\end{eqnarray}
Here we have identified $T_{c0}$ as the critical temperature without impurities. Since $\delta$ is some position from the core of the domain wall with thickness $\approx1/\lambda a$ we find in general 
\begin{eqnarray}
\delta=\frac{1}{\lambda a}+d\delta=\frac{1}{T_c}-\frac{dT_c}{T_c^2}=\frac{1}{T_c}-\frac{bT}{T_c^2},
\label{c5-d}
\end{eqnarray}
where in the last step of Eq.~(\ref{c5-d}) the small contribution ($dT_c$) is associated with a fraction $b$ of some temperature scale $T$. 
Now let us evaluate $T$ at the Kondo scale $T_K$, such that 
\begin{eqnarray}
T\delta&\equiv&T_K\delta=\frac{T_K}{T_c}-\frac{bT_K^2}{T_c^2}\nonumber\\
&=& \frac{T_K}{T_c}\left(1-\frac{bT_K}{T_c} \right)\simeq\frac{T_K}{T_c+bT_K}.
\label{c5-e}
\end{eqnarray}
{Indeed Eq.(\ref{c5-d}) can be seen as few terms of the Taylor expansion of $\delta=1/(T_c+bT)$ in powers of $T$. Alternatively, for an infinite series of such terms Eq.~(\ref{c5-e}) converges to an exact equation.}
The critical temperature (\ref{c5-c}) can be now written as follows
\begin{eqnarray}
\frac{T_c}{T_{c0}} =\exp{\left({-\frac12\frac{\zeta}{\frac{T_c}{T_K}+\frac{T_K}{T_{c0}}}}\right)},
\label{c5-f}
\end{eqnarray}
where we have defined the ratio $b=T_K/T_{c0}$ in order to compare the Kondo scale with respect to $T_{c0}$ --- adopted as an scale of reference. Now for the sake of comparison with previous results in the literature \cite{muller,tami} see Fig.~\ref{figura3a} for $T_K\leq T_{c0}$ (left panel) and $T_K\geq T_{c0}$ (right panel). {Just as a matter of scales, in these plots we have rescaled $\delta\to \delta/\Omega$, such that $\zeta\to\Omega\zeta$ into Eq.~(\ref{c5-f}). In Fig.~\ref{figura3a} we consider $\Omega=30$.}

Now we can explicitly establish the relationship among $\zeta$ and the impurity concentration $\bar{n}$ as follows. By expanding (\ref{c5-f}) in powers of $\zeta$ we find
\begin{eqnarray}
\frac{T_c}{T_{c0}} &=&1-\frac12\frac{\zeta}{\frac{T_c}{T_K}+\frac{T_K}{T_{c0}}}+{\cal O}(\zeta^2)\nonumber\\
 &=&1-\frac12{\frac{T_{c0}}{T_K}}{\zeta}+{\cal O}(\zeta^2), \qquad T_K\gg T_{c0}.
\label{c5-g}
\end{eqnarray}
\begin{figure}[h!]
  \centering
    \includegraphics[width=8.0cm,height=5.5cm]{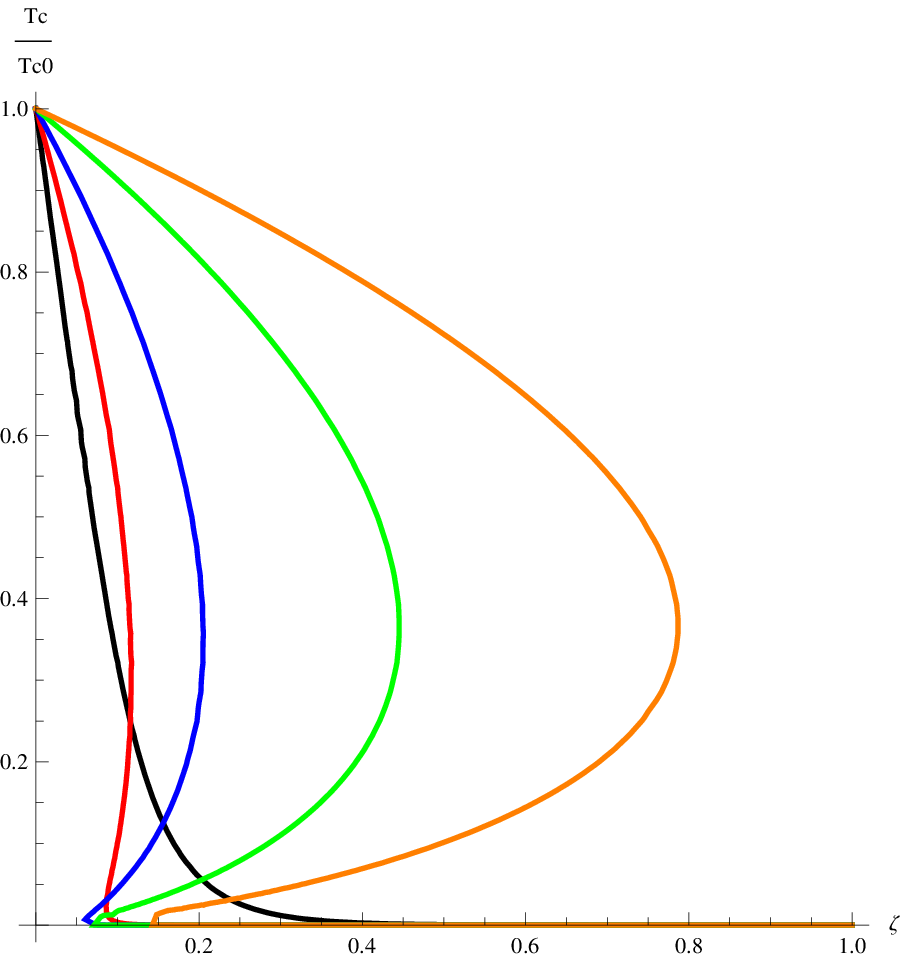}
     \includegraphics[width=8.0cm,height=5.5cm]{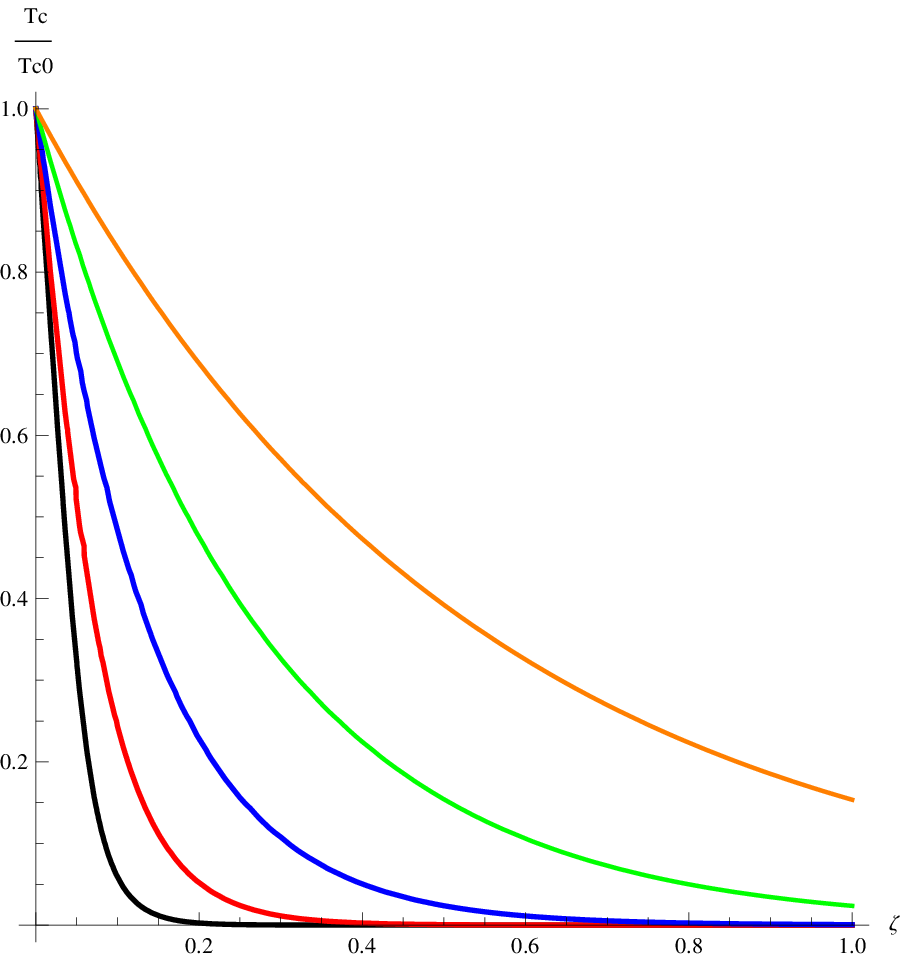}
    \caption{The normalized critical temperature as a function of the Lorentz-violating parameter $\zeta$ (left panel) for $T_K/T_{c0}$ = 1/32, 1/18, 1/8, 1/4 and 1; and $\zeta$ (right panel) for $T_K/T_{c0}$ = 16, 8, 4, 2 and 1 from top to bottom. $T_{c0}=1$.}
    \label{figura3a}
\end{figure}
Then comparing equations (\ref{33c-a}), (\ref{33c-b}) and (\ref{c5-g}) it is easy to establish the correspondence $\zeta\equiv (8\pi^2/c^2)\bar{\eta}$. This makes a clear connection between the Lorentz-violating parameter and impurity concentration in a superconductor, which can display an alternative description of the Kondo effect in quantum field theory.
}

The fact of $T_c$ developing negative curvature for the case $T_K<T_{c0}$ (Fig.~\ref{figura3a} --- left panel) is in full accord with Abrikosov and Gor'kov (AG) theory \cite{AG}. However,  as in the Muller-Hartman and Zittartz (MZ) theory \cite{muller},  our results show a more striking fashion. In this regime we can see that for $T_K$ sufficiently small, $T_c$ becomes no single valued, and at some values of the impurity concentration $\bar{n}$ (or $\zeta$ in our model) one appears three critical temperatures --- this can be easily seen by drawing an imaginary vertical line intercepting three points at some curves. Particularly, there appear two transition temperatures $T_{c1}$ and $T_{c2}$ above $T_K$ for some values of $\bar{n}$. This means that by decreasing the temperature, a system first becomes superconducting at $T_{c1}$ and then becomes normal again at $T_{c2}$. A third critical temperature (if one considers the intercept point at the {\it infinite tails} of the $T_c-\bar{n}$ curves) means that a superconducting phase finally reappears at temperatures below $T_K$, which are hardly accessible experimentally. Thus, this effect showing that superconductivity does not exist at all lower temperatures is an evidence of the Kondo 
effect in superconductors.	 The MZ theory has been shown to be in agreement with several experiments. Particularly, the experimental evidence of three critical temperatures was clearly  found in \cite{Winzer}.

Despite of succeeding in several experiments, the formalism presented in the MZ theory is correct as long as the correlation between impurities can be neglected. The main criticism relies on the fact that the $T_c-\bar{n}$ curve always has an infinite tail, and there appears no critical impurity concentration $\bar{n}_c$ where $T_c$ vanishes. This is in contradiction with experiments that showed there always appears a finite critical concentration --- see \cite{tami} for discussions on extensions of the MZ theory, where, basically, such a critical concentration is achieved by taking into account the dynamical properties of impurities. 

We can see that our model can give such critical impurities if we take into account higher order in $\zeta$ in Eq.~(\ref{c5-f}). An attempt in order to justify these higher order terms from the quantum field theory point of view may be achieved by supplementing the Lorentz-violating term with higher order dynamical Lorentz-violating terms.

\section{Conclusions}
\label{conclu}

In this paper, we extended the domain wall description of superconductivity in a Lorentz-violating model. We have shown that the Lorentz-violating parameter $\zeta$ plays the role of impurities both because it provides a screening of electric charge of the system which affects the optical conductivity and the condensate and also because presents a reduction of the critical temperature in the same way as expected in the Kondo effect. In the Kondo effect such impurities are mainly expected to be magnetic impurities. So this new model may present a more realistic application in describing superconductivity in the realm of quantum field theory in the limit of mean field theory (GL theory) with the inclusion of Lorentz-violating terms.

\acknowledgments
We would like to thank CAPES and CNPq for financial support. The authors also thank Amilcar R. Queiroz for discussions.

\end{document}